\documentclass{emulateapj}
\slugcomment{Accepted for publication by the Astrophysical
Journal} \received{13 December 2006} \accepted{22 March 2007}
\usepackage{graphicx,times}
\def\Journal#1#2#3#4{{#4}, {#1}, {#2}, #3}

\begin{document}
\title{Observation of very high energy $\gamma$-rays from the AGN 1ES\,2344+514 in a
low emission state with the MAGIC telescope}
\shorttitle{Observation of VHE $\gamma$-rays from 1ES\,2344+514}
\shortauthors{J. Albert et~al.}

%
\author{
 J.~Albert\altaffilmark{a},
 E.~Aliu\altaffilmark{b},
 H.~Anderhub\altaffilmark{c},
 P.~Antoranz\altaffilmark{d},
 A.~Armada\altaffilmark{b},
 C.~Baixeras\altaffilmark{e},
 J.~A.~Barrio\altaffilmark{d},
 H.~Bartko\altaffilmark{f},
 D.~Bastieri\altaffilmark{g},
 J.~K.~Becker\altaffilmark{h}
% ...
}
 \altaffiltext{a}{Universit\"at W\"urzburg, D-97074 W\"urzburg, Germany}
 \altaffiltext{b}{Institut de F\'\i sica d'Altes Energies, Edifici Cn., E-08193 Bellaterra (Barcelona), Spain}
 \altaffiltext{c}{ETH Zurich, CH-8093 Switzerland}
 \altaffiltext{d}{Universidad Complutense, E-28040 Madrid, Spain}
 \altaffiltext{e}{Universitat Aut\`onoma de Barcelona, E-08193 Bellaterra, Spain}
 \altaffiltext{f}{Max-Planck-Institut f\"ur Physik, D-80805 M\"unchen, Germany}
 \altaffiltext{g}{Universit\`a di Padova and INFN, I-35131 Padova, Italy}
 \altaffiltext{h}{Universit\"at Dortmund, D-44227 Dortmund, Germany}
% ...

\begin{abstract}

The MAGIC collaboration has observed very high energy gamma ray
emission from the AGN 1ES\,2344+514. A gamma-ray signal
corresponding to an $11\sigma$ excess and an integral flux of
$(2.38\pm{0.30_\mathrm{stat}}\pm{0.70_\mathrm{syst}})\times10^{-11}\,\mathrm{cm}^{-2}\,\mathrm{s}^{-1}$
above 200 GeV has been obtained from 23.1 hours of data taking
between 2005 August 3 and 2006 January 1. The data confirm the
previously detected gamma-ray emission from this object during a
flare seen by the Whipple collaboration in 1995 and the evidence
(below $5 \sigma$ significance level) from long-term observations
conducted by the Whipple and HEGRA groups. The MAGIC observations
show a relatively steep differential photon spectrum that can be
described by a power law with a photon index of $\alpha=-2.95 \pm
0.12_{\mathrm{stat}} \pm 0.2_{\mathrm{syst}}$ between 140 GeV and
5.4 TeV. The observations reveal a low flux state, about six times
below the 1995 flare seen by Whipple and comparable with the
previous Whipple and HEGRA long-term measurements. During the
MAGIC observations no significant time variability was observed.
\end{abstract}


\section{Introduction}

All but one of the detected extragalactic very high energy (VHE)
gamma ($\gamma$) ray sources so far are active galactic nuclei
(AGN) of the BL~Lac type. These objects are characterized by a
highly variable electromagnetic emission ranging from radio to
$\gamma$-rays, and by continuum spectra dominated by non-thermal
emission that consist of two distinct broad components. While the
low energy bump is thought to arise dominantly from synchrotron
emission of electrons, the origin of the high-energy bump is still
debated. Leptonic models ascribe it to inverse Compton processes
that either up-scatter synchrotron photons (synchrotron-self
Compton [SSC] models, Marscher \& Gear 1985, Maraschi et al.
1992), or to external photons that originate from the accretion
disk \citep{DermerSchlickeiser}, from nearby massive stars, or are
reflected into the jet by surrounding material
\citep{SikoraBegelmanRees}.
In hadronic models, interactions of a highly relativistic jet
outflow with ambient matter \citep{DarLaor,Bednarek1993},
proton-induced cascades \citep{PIC}, synchrotron radiation off
protons (proton synchrotron blazar; Aharonian 2000; M\"ucke
\& Protheroe 2001), or curvature radiation, are responsible for
the high energy photons.
The prime scientific interest in BL~Lac objects is twofold: (1) to
understand the VHE $\gamma$-ray production mechanisms, assumed to
be linked to the massive black hole in the center of the AGN, and
(2) to use the VHE $\gamma$-rays as a probe of the extragalactic
background light (EBL) spectrum between about $0.3$
to $30~\mu\rm m$ wavelength. In the past, most of the VHE
$\gamma$-ray emitting AGN were discovered during phases of high
activity, biasing our current observational record towards high
emission states. Although these sources also show variability in
the X-ray, optical, and radio domain, the VHE variability is
observed to often be the most intense and violent one. While
fast variability on the timescale of 10 minutes has been observed
for Mkn~501 in the X-ray domain \citep{XueCui}, flux doubling
times well below 5 minutes were recently also found in the VHE
domain \citep{Gaidos,David}. Many of the observed AGN are
presumably visible only during a state of high activity. It still
remains an open question whether these sources are only
temporarily active and are completely inactive between times of
flaring, or if there also exists a state of low but
continuous $\gamma$-ray emission. In addition, the temporal and
spectral properties of such a low VHE $\gamma$-ray emission state
is mostly elusive as of to date. It is quite conceivable that,
compared to a low state, the flare emission state is either due to
a different population of accelerated particles or originates from
a different region in the AGN, or both.

In the first year of operation of the Major Atmospheric Gamma
Imaging Cerenkov (MAGIC) Telescope a program has been started to
search for new low and medium redshift blazars emitting at VHE
$\gamma$-rays (Albert et al. 2006b; 2006c; 2007a). In addition, known VHE AGNs
were monitored in order to study common features of their
$\gamma$-ray emission, as well as the properties of the low-emission state (Albert et al. 2006a; 2007b; 2007c).

A good candidate for detailed studies is 1ES\,2344+514. This AGN
belongs to a type of blazars in which the synchrotron emission
peaks at UV/X-ray frequencies (the so-called high-energy peak BL
Lacs [HBLs] e.g. Urry \& Padovani 1995), as opposed to the blazars
with the synchrotron peak located at IR/visible frequencies.
Along with Mkn\,501 and H\,1426+428, it represents extreme BL Lac
objects, in which the synchrotron peak energy exceeds 10~keV, in
particular during strong flares \citep{Costamante}. 1ES\,2344+514
was detected during the {\it Einstein} Slew Survey \citep{Elvis}
in the energy range between 0.2 and 4 keV. It was identified as a
BL~Lac object by \citet{Perlman}, who also determined a redshift
of $z = 0.044$. Its black hole mass was estimated to be
$10^{(8.80\pm0.16)} M_\mathrm{sun}$ \citep{Barth}. Early {\it
BeppoSAX} observations \citep{Giommi} revealed a large 0.1$-$10
keV flux variability on timescales of a few hours. Follow-up
observations in 1998 found the object in a very low state with the
synchrotron peak shifted by a factor of 20 towards lower energies
and the corresponding integral flux decreased by a factor of 4.5.
\citet{Giommi} interpreted the observations with one electron
population being responsible for the steady low energy synchrotron
emission and another electron component producing higher energy
X-rays with high time variability. The latter component should be
responsible for VHE $\gamma$-ray emission via inverse Compton (IC)
scattering. EGRET did not detect any signal from 1ES\,2344+514,
giving an upper limit of
$3.4\times10^{-11}\,\mathrm{erg}\,\mathrm{cm}^{-2}\,\mathrm{s}^{-1}$
at its peak response energy of 300~MeV \citep{Fichtel}. During the
winter of 1995/1996, the Whipple collaboration reported a $5.8\sigma$
excess signal from 1ES\,2344+514 above 350 GeV from 20.5 hours
observation time \citep{Catanese}. The observed flux was highly
variable, with the most significant signal occurring during a
flare on 1995 December 20, while all the remaining data combined
led to an only marginal ($4\sigma$) excess, i.e. below the
canonical detection limit used in ground-based VHE $\gamma$-ray
astronomy. The 0.8$-$12.6 TeV differential spectrum measured by
the Whipple collaboration during the flare had a power-law index
of $-2.54\pm{0.17_\mathrm{stat}}\pm0.07_{\mathrm{syst}}$
\citep{Schroedter}. One year later another search did not reveal
any VHE $\gamma$-ray emission. The HEGRA collaboration also
searched for VHE $\gamma$-ray emission above 800 GeV. A deep
exposure of 72.5~h indicated a signal at a significance level of
$4.4\sigma$ \citep{Tluczykont}.

Here we present MAGIC telescope observations of 1ES\,2344+514. We
briefly discuss the observational technique used and the
implemented data analysis procedure, derive a VHE $\gamma$-ray
spectrum of the source, and put the results into perspective with
other VHE $\gamma$-ray observations of this AGN. An SSC model is
used to describe the wide-range spectral energy distribution
(SED).

\section{Observations}
The observations were performed between 2005 August 3 and 2005
September 29, and between 2005 November 11 and 2006 January 1,
using the MAGIC Telescope on the Canary island of La Palma
($28.8^\circ$N, $17.8^\circ$W, 2200~m above sea level), from where
1ES\,2344+514 can be observed at zenith distances above
$24^\circ$. The essential parameters of the currently largest air
Cherenkov telescope are a 17~m~\o\ segmented mirror of parabolic
shape, an $f/D$ of 1.05 and a hexagonally shaped camera with a
field of view (FOV) of $\approx3.5^\circ$ mean diameter. The
camera comprises 576 pixels composed of hemispherical, six dynode
photomultipliers augmented in sensitivity by a diffuse lacquer
doped with a wavelength shifter \citep{Paneque} and by so-called
light catchers. In separate measurements a total gain of 2 has
been determined. 180 pixels of $0.2^\circ$~\o\ surround the inner
section of the camera, which consists of 394 pixels of
$0.1^\circ$~\o\ ($= 2.2^\circ$~\o\ FOV). The trigger is formed by
a coincidence of $\geq 4$ neighboring pixels.
The overall Cherenkov photon ($300-650$~nm) to photoelectron
conversion ratio is $0.15\pm0.02$. 
The point spread function (PSF) of the main mirror is
$\sigma\approx 0.04^\circ$, while 90\% of the light of a source at
infinity is focussed onto a disk with $0.1^\circ$~\o. Further
details of the telescope parameters and performance can be found
in \citet{MAGIC-commissioning,CortinaICRC}.

1ES\,2344+514 was observed for 32 hours in total, distributed over
27 days between 2005 August and the first days of 2006 January at
zenith angles ranging from $23^\circ$ to $38^\circ$. The
observations were carried out in wobble mode \citep{Fomin}, i.e.
by alternatingly tracking two positions at $0.4^\circ$ offset from
the camera center. This observation mode allows a reliable
background estimation for point sources.

Simultaneous $R$-band observations of 1ES\,2344+514 were conducted
in the framework of the Tuorla Observatory blazar monitoring
program\footnote{See http://users.utu.fi/kani/1m/.} with the KVA
35~cm telescope\footnote{See http://tur3.tur.iac.es/.} on La Palma
and the 1.03~m telescope at Tuorla Observatory, Finland.

\section{Data Analysis}

The data analysis was carried out using the standard MAGIC
analysis and reconstruction software \citep{Magic-software}. After
calibration \citep{MAGIC_calibration}, the images were cleaned by
requiring a minimum number of seven photoelectrons (core pixels) and five
photoelectrons (boundary pixels), see e.g. \citet{Fegan1997}.
These tail cuts are scaled accordingly for the larger size of the
outer pixels of the MAGIC camera. The data were filtered by
rejecting trivial background events, such as accidental noise
triggers, triggers from nearby muons or data taken
during adverse conditions (low atmospheric transmission, car light
flashes etc.). Light clusters, either from large angle shower
particles or from the night sky light background (stars), well
separated from the main image, were removed from the images. For
the events included in the analysis, the mean trigger rate was
required to be constant within $\approx 20\%$. In order to improve
the comparability of the two data sets from summer and winter 2005
we restricted the maximum zenith angle to $\leq 34^\circ$.
From the remaining events, corresponding to 23.1 h observation
time, image parameters were calculated \citep{Hillas_parameters}
such as WIDTH, LENGTH, SIZE, CONC, and M3LONG, the third moment of
the light distribution along the major image axis. For the
$\gamma$/hadron separation a multidimensional classification
procedure based on the random forest method was employed
\citep{Breiman,Bock}. The separation procedure was trained using a
sample of Monte Carlo (MC) generated $\gamma$-ray shower images
\citep{Knapp,Majumdar} on the one hand and about 1\% randomly
selected events from the measured wobble data representing the
hadronic background on the other hand. The MC $\gamma$-ray showers
were generated between zenith angles of 24$^\circ$ and 34$^\circ$
with energies between 10~GeV and 30~TeV. Every event was assigned
a parameter called hadronness ($h$), which is a measure for the
probability that it is a hadronic (background) event. The final
separation was achieved by a cut in $h$. The same cut procedure
was applied to the final 1ES\,2344+514 sample. The arrival
directions of the showers in equatorial coordinates were
calculated using the DISP method
\citep{Fomin,Lessard2001,MAGIC_disp}. The energy of the primary
$\gamma$-ray was reconstructed from the image parameters again
using the random forest method and taking into account the full
instrumental energy resolution.

\begin{figure}[!h]
\begin{center}
\includegraphics[width=0.75\linewidth]{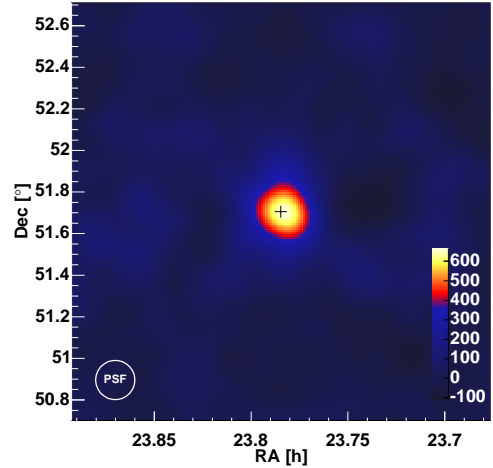}
\end{center}
\caption{Sky map for 1ES\,2344+514 produced with a DISP analysis:
The figure shows the (background-subtracted, see e.g. Rowell 2003)
excess events above 300 photoelectrons, corresponding to a
$\gamma$-ray energy of $\approx 180$ GeV.
The sky map has been smoothed using a two dimensional Gaussian
with $\sigma=0.1^\circ$, roughly corresponding to the $\gamma$ PSF
of the MAGIC telescope for point sources (indicated by the white
circle). The colors encode the number of excess events in units of
$10^{-5} \mathrm{sr}^{-1}$. The black cross marks the expected
source position.} \label{fig:skymap}
\end{figure}

\begin{figure}[!h]
\begin{center}
\includegraphics[width=0.95\linewidth]{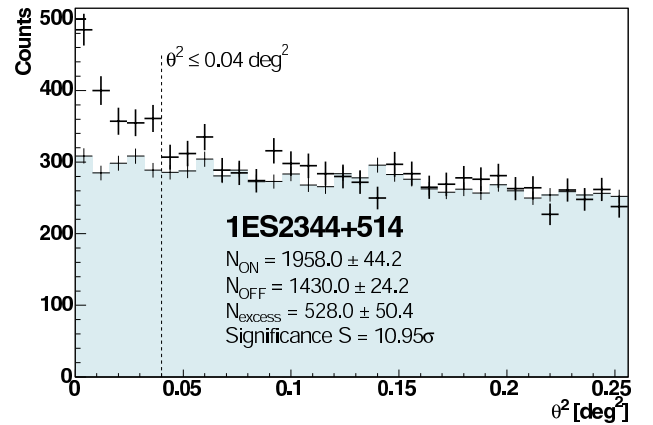}
\end{center}
\caption{$\theta^2$ plot for the 1ES\,2344+514 observations.
On-source events are given as black symbols, while red symbols
represent off-source background. A cut at $\theta^2 \leq
0.04\,\mathrm{deg}^2$ selects 528 $\gamma$ events at a
significance level of $11\sigma$. The plot has been prepared for
SIZE~$>300$ photoelectrons, corresponding to $\approx 180$ GeV.}
\label{fig:t2}
\end{figure}

Fig.~\ref{fig:skymap} shows a $\approx 1.8^\circ \times 1.8^\circ$
section of the sky around the 1ES\,2344+514 position. The nominal
source position is marked by a cross. A clear excess is visible in
the data, the maximum of which is located at
(RA,Dec)=$(23^{\mathrm h}46^{\mathrm m}\pm 0\fm4,
51^{\circ}42\farcm6 \pm 1\farcm2)$ (the errors only include the
determination accuracy of the position). The extension of the
excess and the small deviation from the nominal position are
consistent with the PSF and the tracking error of $\approx 1\fm5$
of the telescope \citep{MAGIC_drive}, respectively. To calculate
the significance of the observed $\gamma$-ray excess, the squared
angular distance $\theta^2$ between the reconstructed shower
direction and the object position ($\theta^2=0$) as shown in
Fig.~\ref{fig:t2} is used. In this representation, the background
is expected to be flat for the case of a very large diameter
camera. In the analysis, three background regions of the same size
chosen symmetrically with the source position around the camera
center were used for a simultaneous determination of the
background. The background control data sample was normalized to
the on-source sample between
$0.12\,\mathrm{deg}^2<\theta^2<0.24\,\mathrm{deg}^2$. The
reason of the slow but steady drop in the background is the drop
in acceptance towards the camera boundary.
The observed excess signal 
of 528 events below $\theta^2<0.04\,\mathrm{deg}^2$ corresponds to
an $11\sigma$ excess according to eq. 17 in \citet{LiMa}. An
independent analysis using other cuts, a different reconstruction
algorithm and a different $\gamma$/hadron optimization procedure,
revealed a comparable (within statistics) significance. While for
the sky map and the $\theta^2$ plot a fixed, tight $h$ cut was
applied, the final separation for the spectral analysis and the
light curve was done using a looser, energy-dependent cut in $h$,
requiring that about 60\% of the MC $\gamma$ events survive.

\begin{figure}
\includegraphics[width=\linewidth]{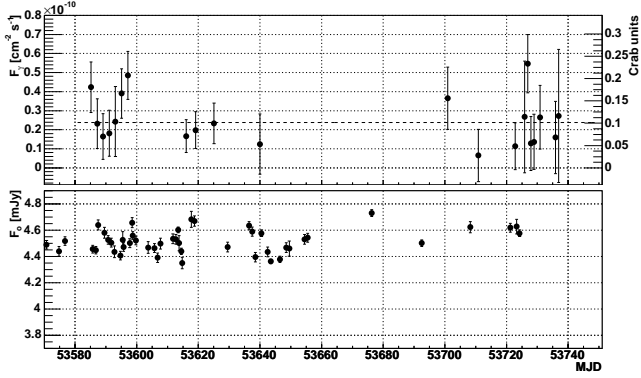} \caption{VHE ($E>200
\mathrm{GeV}$) light curve (upper panel) and simultaneous optical
($R$-band) light curve for 1ES\,2344+514. The dashed line in the
VHE light curve indicates the average flux level of
$(2.38\pm0.30)\times10^{-11} \mathrm{cm}^{-2} \mathrm{s}^{-1}$
($\chi^2_\mathrm{red}=21.2/20$). Note that the contribution of the
host galaxy to the optical brightness is non-negligible and given
as 3.7 mJy \citep{KariO}.} \label{fig:lc}
\end{figure}

As the analyzed data comprise 21 observation nights, it is
possible to check the light curve for possible flux variability.
On a diurnal basis, the $\geq$~200~GeV light curve
(Fig.~\ref{fig:lc}) shows small changes and trends beyond those
expected from statistical fluctuations. The structure observed
during MJD 53580$-$53600 is compatible with a constant-flux ansatz
($\chi^2/$dof$=6.1/6$), while from MJD 53726.82$-$53726.90 a flux
of $2.4\sigma$ above the average flux inferred from the
surrounding days MJD 53720$-$53740, $(1.8\pm0.6)\cdot10^{-11}
\mathrm{cm}^{-2} \mathrm{s}^{-1}$ ($\chi^2/$dof=4.9/7), was found.
Note that the probability for finding such an excess in the 21
observation nights is around $34.4\%$. No significant variability
within this observation night, encompassing 1.13 hours of
effective observation time, can be claimed.

The observation time can be split into three observation periods
(Table~\ref{tab:lc}). Together with the VHE $\gamma$-ray  light
curve, an $R$-band optical light curve is shown. Simultaneous
X-ray data are only available from the ASM
instrument\footnote{Data available at http://xte.mit.edu/.}
\citep{Levine} on board the {\it Rossi X-ray Timing Explorer}, the
sensitivity of which, however, would be hardly sufficient to
resolve the expected $2-10$ keV flux even during flaring states of
1ES\,2344+514, like those observed with {\it BeppoSAX}
\citep{Giommi}.

\begin{table}
\begin{center}
\begin{tabular}{llcc}
\hline
Period         & Obs. time         & $F_{>200\mathrm{GeV}}$ & $\chi^2/$ndf \\
(MJD)          &                   & ($10^{-11}\,\mathrm{cm}^{-2}\,\mathrm{s}^{-1}$)  &  \\
\hline
$53585-53597 $ & \phantom{0}6.37~h & $3.02\pm0.50$ & $6.1/6$ \\
$53610-53642 $ & \phantom{0}8.06~h &  $1.87\pm0.52$ & $0.4/3$ \\
$53700-53736 $ & \phantom{0}8.66~h & $2.20\pm0.51$ & $12.0/9$ \\
\hline
Combined       &           23.09~h &  $2.38\pm0.30$ & $21.2/20$ \\
\hline
\end{tabular}
\end{center}
\caption{Integral fluxes above 200 GeV in the individual
observation periods and reduced $\chi^2$/dof of a fit with a
constant-flux ansatz in the respective observation periods. The
given errors are statistical errors only.} \label{tab:lc}
\end{table}

Summing up all the data we determined an integral flux above 200
GeV of
\begin{displaymath}
F(E>200 \mathrm{GeV}) =  (2.38 \pm {0.30_{\mathrm{stat}}} \pm
0.70_{\mathrm{syst}}) \times 10^{-11} \mathrm{cm}^{-2}
\mathrm{s}^{-1}.
\end{displaymath}
The relatively large systematic error is a consequence of the
steep spectral slope (see below). The main contributions to the
systematic error are the uncertainties in the atmospheric
transmission, the reflectivity (including stray-light losses) of
the main mirror and the light catchers, the photon to
photoelectron conversion calibration, and the photoelectron
collection efficiency in the photomultiplier front end. Also, MC
uncertainties and systematic errors from the analysis methods
contribute significantly to the error. The above quoted flux
corresponds to $(10\pm1)\%$ of the integral Crab Nebula flux in
the same energy range.
%

During our observations we also checked the optical variability.
When correcting for the contribution of the host galaxy of 3.7~mJy
\citep{KariO}, variations in the optical light curve around the
average brightness of $\approx 15\%$ are seen, which are
significant given the small errors ($\lesssim 5\%$) of the data
points. Possible VHE $\gamma$-ray variations on a comparable level
are below the sensitivity of MAGIC on the given timescale.

\begin{figure}[h]
\begin{center}
\includegraphics[width=\linewidth]{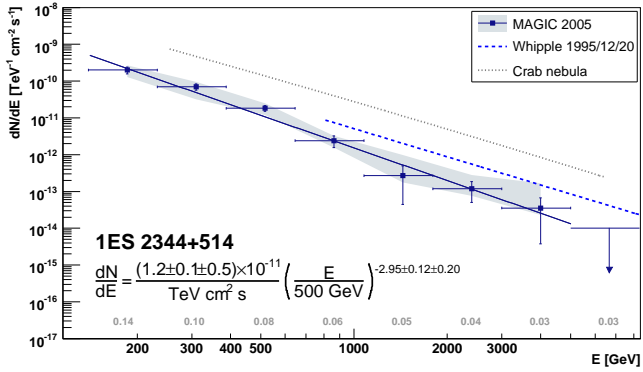}
\end{center}
\caption{Differential photon spectrum for 1ES\,2344+514 as
measured with MAGIC (solid curve). The gray band represents
systematic errors coming from varying the $\gamma$ efficiency in
the determination of the spectrum. The Crab nebula spectrum as
measured with MAGIC is also shown (gray dotted curve; small gray
numbers indicate the fraction of Crab nebula flux for the
1ES\,2344+514 flux points) and the Whipple flare spectrum (dashed
curve) as reported by \citet{Schroedter}.} \label{fig:spectrum}
\end{figure}

For each of the three observation periods photon spectra were
determined. These are well described by simple power laws between
140 GeV and at least 1.0 TeV and are, within errors, compatible
with no change in the spectral index.
Finally, all data were combined for the calculation of a
differential photon spectrum (Table \ref{tab:spd}). The
reconstructed spectrum after unfolding with the instrumental
energy resolution \citep{Anykeev1991,Mizobuchi} is shown in
Figure~\ref{fig:spectrum}. A simple power law fit to the data
between 140 GeV and 5.4 TeV yields
\begin{displaymath}
\frac{\mathrm{d}N}{\mathrm{d}E} =
\frac{(1.2\pm0.1_{\mathrm{stat}}\pm{0.5_\mathrm{syst}}) \cdot
10^{-11}} {\mathrm{TeV}\,\mathrm{cm}^2\,\mathrm{s}} \frac{E}{500\,
\mathrm{GeV}}^{-2.95\pm{0.12_\mathrm{stat}}\pm{0.2_\mathrm{syst}}}
\end{displaymath}
with a reduced $\chi^2$/dof of 8.56/5, indicating a
reasonable description of the data by the fit. For comparison, the
Whipple measurement of the 1ES\,2344+514 spectrum during the flare
of 1995 December 20 \citep{Schroedter} and the Crab Nebula spectrum
\citep{Crab_MAGIC} as obtained with MAGIC are also shown in
Fig.~\ref{fig:spectrum}. Note that the integral flux
$F(E>970\,\mathrm{GeV})=(0.82\pm0.09)
\times10^{-12}\,\mathrm{cm}^{-2}\,\mathrm{s}^{-1}$ is in very good
agreement with the HEGRA measurements from 1998 to 2002
\citep{Tluczykont}.

\begin{table}
\begin{center}
\begin{tabular}{cccccc}
\hline
Mean energy & Bin Width & Flux & Stat. Error & \multicolumn{2}{c}{Syst. Error} \\
$E$ [GeV]   & [GeV]     & \multicolumn{4}{c}{[TeV$^{-1}$ cm$^{-2}$ s$^{-1}$]} \\
\hline
\phantom{0}186 & \phantom{00}93 & $\phantom{<}2.0$E-10 & 4.2E-11 & +7.0E-11 & -7.0E-11 \\
\phantom{0}310 & \phantom{0}155 & $\phantom{<}7.0$E-11 & 1.4E-11 & +3.8E-11 & -2.4E-11 \\
\phantom{0}516 & \phantom{0}259 & $\phantom{<}1.8$E-11 & 3.2E-12 & +6.4E-12 & -6.4E-12 \\
\phantom{0}861 & \phantom{0}431 & $\phantom{<}2.4$E-12 & 8.6E-13 & +8.4E-13 & -8.4E-13 \\
          1437 & \phantom{0}720 & $\phantom{<}2.7$E-13 & 2.2E-13 & +9.4E-14 & -7.1E-13 \\
          2397 &           1201 & $\phantom{<}1.2$E-13 & 6.8E-14 & +4.1E-14 & -1.6E-13 \\
          3999 &           2003 & $\phantom{<}3.5$E-14 & 3.2E-14 & +1.2E-14 & -1.3E-13 \\
          6670 &           3341 &           $<8.4$E-15 & \multicolumn{3}{l}{(95\% C. L.)} \\
\hline
\end{tabular}
\end{center}
\caption{Differential flux of 1ES\,2344+514 along with statistical
and systematical errors.} \label{tab:spd}
\end{table}

\begin{figure*}
\includegraphics[width=\linewidth]{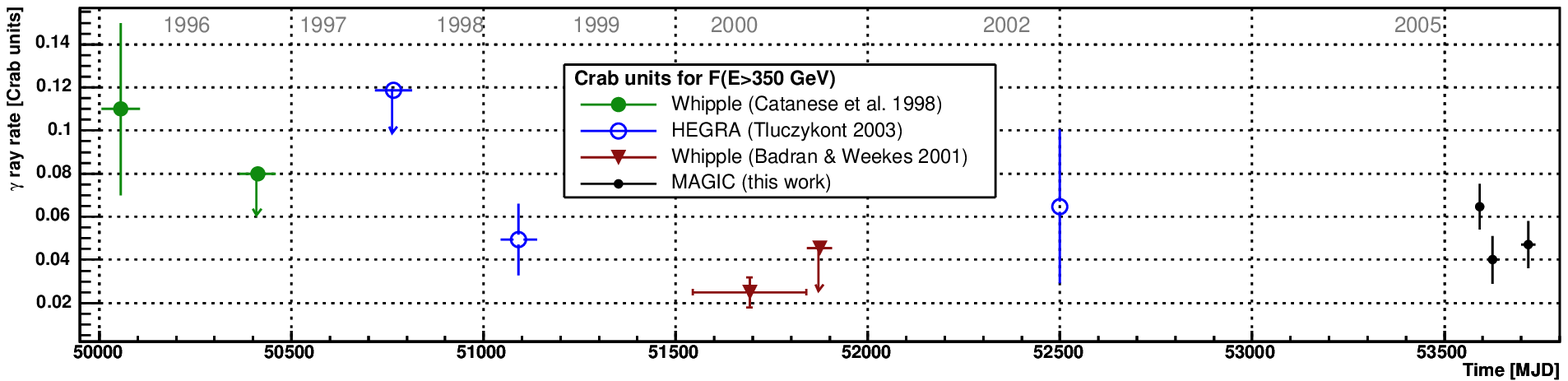}
\caption{Overall VHE light curve for 1ES\,2344+514. Data for
1996-2002 were collected from
\citet{Catanese,2001ICRC....7.2892B,Tluczykont}. The 1995 December
20 flare has been excluded for clarity. Due to the different
energy thresholds of the included observations, integral fluxes
$\geq 350$~GeV were considered and the data were normalized to
Crab flux units, extrapolating data points with higher flux levels
assuming 1ES\,2344+514 was observed in a quiescent state; the
spectral behavior found in this paper was used for this purpose.}
\label{fig:lccom}
\end{figure*}

\section{Discussion}

We observed a highly significant VHE $\gamma$-ray emission from
the blazar 1ES\,2344+514. 
The flux exhibited no significant variations on a timescale of
days, with one night showing a higher flux level (by a factor of
2) as compared to the surrounding nights; such a variation is
statistically expected to occur with a 34.4\% probability in 21
observation nights. The observed flux was lower by a factor of
$\approx6$ than the one observed by the Whipple collaboration
during a flare on 1995 December 20. The spectrum is softer than
the Crab Nebula spectrum and also softer than the flare spectrum
observed by the Whipple collaboration.

\subsection{Long-term VHE Light Curve}

In Fig.~\ref{fig:lccom} we show a light curve including all
reported VHE $\gamma$-ray measurements and upper limits for
1ES\,2344+514. These data have been normalized to an integral Crab
flux $F(E>350\,\mathrm{GeV})$; the fluxes given in the literature
were extrapolated, if necessary, using the spectral index found in
this paper (in the following, ``Crab units'' refer to this energy
threshold). Except for the 1995 December 20 flare and the MAGIC
data, all these measurements are on the sensitivity level of the
respective instruments. Therefore, none of the latter data points
exceeds a significance of $4.3\sigma$. All reported observations
with significances below $2.0\sigma$ ($\approx$95\% probability)
were converted to $99\%$ upper flux limits.

In 1995/1996, Whipple discovered 1ES\,2344+514 at a flux level of
$(0.11\pm0.05)$ Crab units at $E>350$~GeV, except for the 1995 December
flare, when $(0.63\pm0.15)$ Crab units were obtained
\citep{Catanese}. Follow-up observations by Whipple and HEGRA in
1996-1998 yielded upper limits of $0.08$ and $0.12$ Crab
units, respectively. In 1998 and 2002, the object was observed for
almost 60~h by HEGRA resulting, when combined, in a flux of
$(0.042\pm0.012)$ Crab units at $E>930$~GeV \citep{MartinPhd},
which translates to $(0.053\pm0.015)$ Crab units when
extrapolating to $E\geq350$~GeV. From observations of
1ES\,2344+514 in 2002, the Whipple group could infer a low flux
level of $\lesssim0.03$ Crab units with a marginal significance of
$3.1\sigma$ \citep{2001ICRC....7.2892B} at $E\gtrsim400$~GeV.

While the Whipple and HEGRA measurements allowed to conclude on a
emission level of $\leq 11\%$ Crab units only after long
observation times, the MAGIC observations obtained in this paper
are the first time-resolved measurements at this emission
level for 1ES\,2344+514. We find the flux of 1ES\,2344+514 to be
$(0.054\pm0.006)$ Crab units for $E>350$~GeV, which is well in
line with the HEGRA 1997-2002 evidence.

In previous observations of 1ES\,2344+514 it was not possible
to infer temporal characteristics of the found VHE $\gamma$-ray
emission level. With MAGIC, this level can be detected with only a
few hours of observations, enabling studies of the VHE
$\gamma$-ray variability properties of this object over a
significant part of its dynamical range. Thus, 1ES\,2344+514 adds
to the small group of blazars for which such studies are now
possible on a diurnal basis---Mkn 421 \citep{Daniel}, Mkn 501
\citep{David}, and PKS 2155-304 \citep{2155}.

\subsection{Intrinsic Energy Spectrum}
\begin{figure}
\begin{center}
\includegraphics[width=\linewidth]{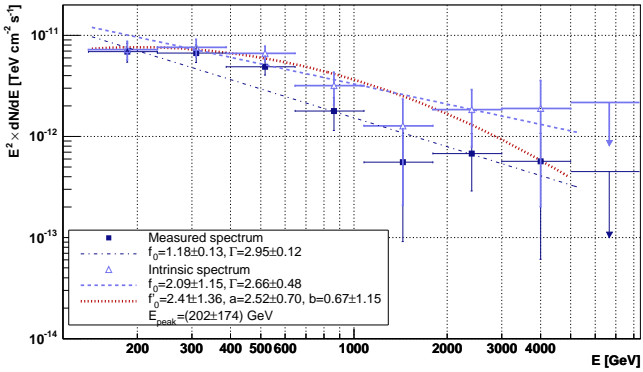}
\end{center}
\caption{Measured and intrinsic differential photon spectrum of
1ES\,2344+514. Simple power law fits are given by the dot-dashed
and dashed lines, while the fit with a logarithmic curvature term
is given by the dotted curve.} \label{fig:spectrumdeabs}
\end{figure}

Having to traverse a cosmological distance corresponding to a
redshift of $z=0.044$, the $\gamma$-rays emitted by 1ES\,2344+514
interact with the low energy photons of the EBL (see, e.g., Nikishov 1962;
Gould \& Schr\'eder 1966; Hauser  \& Dwek 2001). The predominant reaction $
\gamma_{\mbox{\scriptsize{\,VHE}}} + \gamma_{\mbox{\scriptsize
EBL}} \rightarrow \mathrm{e}^{+}\,\mathrm{e}^{-}$ leads to an
attenuation of the intrinsic spectrum
$\mathrm{d}N/\mathrm{d}E_\mathrm{intr}$ that can be described by
\begin{equation}
\mathrm{d}N/\mathrm{d}E_\mathrm{obs} \,=\,
\mathrm{d}N/\mathrm{d}E_\mathrm{intr} \cdot
\exp[-\tau_{\gamma\gamma}(E,\,z)]
\end{equation}
with the observed spectrum $\mathrm{d}N/\mathrm{d}E_\mathrm{obs}$,
and the energy-dependent optical depth
$\tau_{\gamma\gamma}(E,\,z)$.
Here we use the ``best-fit'' model of \citet{kneiske}, which
yields an EBL spectrum that agrees with alternative models, e.g.
\citet{stecker}.
Using this EBL spectrum and a state-of-the-art cosmology (flat
universe, Hubble constant $H_0=72\,\mathrm{km}\,\mathrm{s}^{-1}\,
\mathrm{Mpc}^{-1}$, matter density $\Omega_\mathrm{m}=0.3$, and dark
energy density $\Omega_{\Lambda}=0.7$), we calculate the optical
depth $\tau_{\gamma\gamma}$ for the distance of 1ES\,2344+514.
Thereby we use the numerical integration given by eq.~2 in
\citet{dwek}.
The reconstructed intrinsic source spectrum is shown along with
the measured spectrum in Fig.~\ref{fig:spectrumdeabs}. The
intrinsic source spectrum can be described by a simple power law
of the form
\begin{displaymath}
\frac{\mathrm{d}N}{\mathrm{d}E}_\mathrm{intr} = \frac{
(2.1\pm{1.2_\mathrm{stat}}\pm{0.5_\mathrm{syst}})\cdot10^{-11}}{
\mathrm{TeV}\,\mathrm{cm}^2\,\mathrm{s}} \times \frac{E}{500\,
\mathrm{GeV}}^{-2.66\pm{0.50_\mathrm{stat}}\pm{0.20_\mathrm{syst}}}
\end{displaymath}
between 140 GeV and 5.4 TeV ($\chi^2_{\mathrm{dof}} = 0.68/5$).
The spectrum shows a tendency to flatten towards low energies. A
fit with a logarithmic curvature term, which corresponds to a
parabolic law in a $\log(E^2 \mathrm{d}N/\mathrm{d}E)$ vs. $\log
E$ representation (power law index $\alpha \rightarrow a + 2b \log
(E/E_a)$; Massaro et al. 2004), shows a clear curvature and
enables locating the peak at $E_{\rm peak} = E_a \cdot
10^{(2-a)/(2b)}=(202\pm174)$ GeV. The peak is obviously badly
determined as the turnover of the spectrum, presumably around
$200$ GeV, is not observed unambiguously.

\subsection{Spectral Energy Distribution}

\begin{figure*}
\begin{center}
\includegraphics[width=\linewidth]{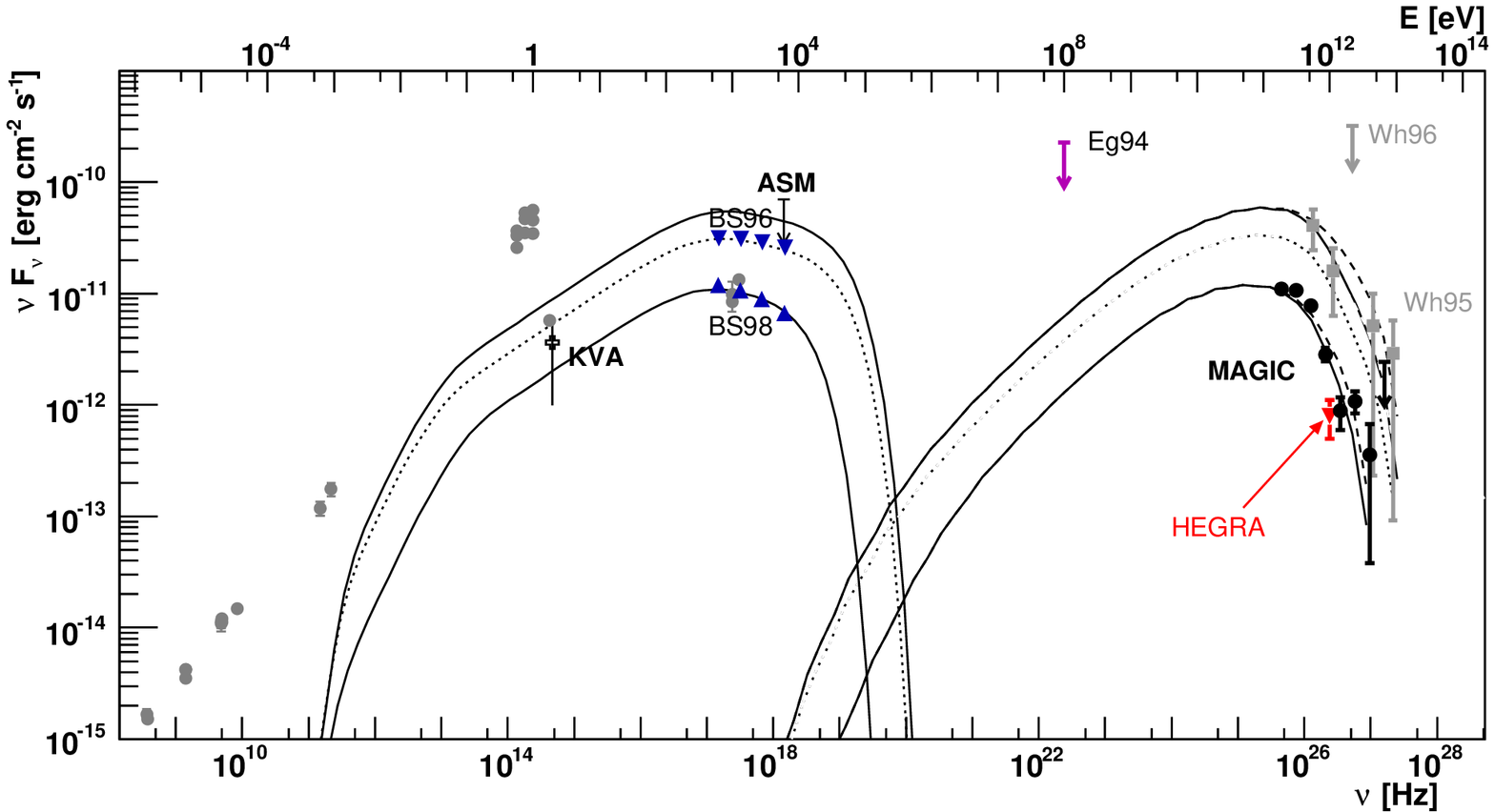}
\end{center}
\caption{Overall SED for 1ES\,2344+514. Gray symbols: Archival
(radio, optical, X-ray) data taken from \citet{BSASDC,Schroedter}.
The two {\it BeppoSAX} data sets represent a quiescent state and
data taken simultaneously with Whipple observations: BS96---{\it
BeppoSAX} 1996 December 05; BS98---{\it BeppoSAX} 1998 June 28.
Wh95---Whipple flare spectrum; Wh96---Whipple upper limit
corresponding to the BS96 measurement \citep{Schroedter}.
Eg94---{\it EGRET} upper limit \citep{hartman}. HEGRA 1998-2002
flux point \citep{Tluczykont}; MAGIC---this paper; data taken
simultaneously with the MAGIC measurements: KVA: Optical flux,
host galaxy contribution subtracted; ASM: RXTE--ASM upper limit.
The solid curves were obtained using the model given in
\citet{Krawczynski1es1959-2004} and describe the synchrotron and
IC emission. The corresponding intrinsic (EBL de-absorbed) spectra
are indicated by the dashed curves. The solid lines model the
flare state of 1995 and the low state as seen by MAGIC in 2005.
The dotted curve is to describe the BS96/Wh96 observation and only
differs in a lower Doppler factor ($\delta=13.2$) from the Whipple
flare model. } \label{fig:wsed}
\end{figure*}

The spectral energy distribution for 1ES\,2344+514 is shown in
Fig.~\ref{fig:wsed}. Apart from the VHE $\gamma$-ray observations
by MAGIC, Whipple, and HEGRA, the X-ray measurements performed by
{\it Einstein}, {\it ROSAT}, and {\it BeppoSAX} and an upper
limit from {\it EGRET} are the most relevant measurements for
modeling the SED. The figure also includes optical and radio
measurements. The latter however cannot be described by the
homogeneous one-zone SSC model provided by
\citet{Krawczynski1es1959-2004} that was used here, as the radio emission
is thought to arise from a larger volume in the jet than the VHE $\gamma$-ray
emission and as self-absorption effects are not accounted for.

Two {\it BeppoSAX} data sets are shown: one was obtained
simultaneously with Whipple observations on 1996 December 5. The
second set was taken during a quiescent period of 1ES\,2344+514,
and represents a rather low X-ray emission state of the object,
although it might not necessarily correspond to its state during
the MAGIC measurements. It should be emphasized that most of the
data points shown in the SED were not measured simultaneously,
which makes the SED modeling very difficult.
The input parameters of the homogeneous SSC model are the radius
of the spherically assumed emission region $R$, the Doppler factor
$\delta$, the magnetic field strength $B$ in the acceleration
region, the density $\rho$ of the electrons responsible for the
$\gamma$-ray emission as well as two spectral slopes and a
spectral cutoff of the electron spectrum.
%
%
The size of the emission region $R$ was chosen such that it can
account for day-scale variability along with the Doppler factor
$\delta$ chosen, $R \leq \delta_{-1} \times t_{\mathrm{d}} \times
2.48 \times 10^{16} \mathrm{cm}$ with $\delta_{-1}=\delta / 10$ and
$t_\mathrm{d}$ in units of days.
The parameters used here are specified in Table~\ref{tab:wsedp}.

\begin{table}
\begin{center}
\begin{tabular}{lll}
\hline
SSC model input & Low state & Flare SED \\
parameter       & SED (MAGIC)   &  \\
\hline
Doppler factor $\delta$ & 8.4 & 15.2 \\
Magnetic field strength $B$ & 0.095 G & 0.075 G \\
Emission region radius $R$ & $10^{16}$~cm & $10^{16}$~cm \\
Electron density $\rho$ & 0.025 erg cm$^{-3}$ & 0.025 erg cm$^{-3}$ \\
${E_{\mathrm{min}}} [\log E({\rm eV})]$ & 9.1 & 8.9 \\
${E_{\mathrm{max}}} [\log E({\rm eV})]$ & 11.6 & 11.9 \\
${E_{\mathrm{br}}} [\log E({\rm eV})]$ & 10.9 & 10.9 \\
$n(E_\mathrm{min}<E\leq {E_{\mathrm{br}}})$ & -2.2 & -2.2 \\
$n(E_\mathrm{br}<E\leq {E_{\mathrm{max}}})$ & -3.2 & -3.2 \\
\hline
\end{tabular}
\caption{Model input parameters used with the SSC model provided
by \citet{Krawczynski1es1959-2004} for describing the MAGIC low
emission state and the 1995 Whipple flare emission state as
depicted in Fig.~\ref{fig:wsed}.} \label{tab:wsedp}
\end{center}
\end{table}

The two models shown are to represent both the flare state of 1995
and the MAGIC observations in 2005. They differ in the following
model input parameters: (1) in the Doppler factor, (2) in the
magnetic field strength, and (3) the minimum and maximum electron
energy. In addition, the form of the electron spectrum differs for
the two flux states. The radius of the emission region was kept
constant at $10^{14}$~m.
While most of the model parameters are compatible with the
parameter space spanned by other models for blazars (e.g. Kino et
al. 2002; Giommi et al. 2002), the magnetic field strength found
here is rather low.

In conclusion, we note that the presented SED models are rather
speculative, given the non-simultaneity of the currently available
data. Future multiwavelength campaigns on 1ES\,2344+514,
exploiting the enhanced sensitivity of the new imaging air
Cherenkov telescope installations, will hopefully improve this
situation.

\acknowledgements

We thank the IAC for the excellent working conditions at the
Observatorio del Roque de los Muchachos in La Palma. The support
of the German BMBF and MPG, the Italian INFN and the Spanish CICYT
is gratefully acknowledged. This work was also supported by ETH
Research Grant TH~34/04~3 and the Polish MNiI Grant 1P03D01028.

\clearpage

\clearpage

\clearpage

\clearpage

\clearpage

\end{document}